\documentclass{aa}  

\usepackage{graphicx}
\usepackage[dvipsnames]{xcolor}
\usepackage{txfonts}
\newcommand{\msun}{M$_{\rm \odot}$}
\newcommand{\vsini}{$V\sin i$}

\begin{document}

   \title{Testing the Role of Merging Binaries in the Formation of the Split Main Sequence in Young Clusters}

\titlerunning{Split MS and Binaries}
   \author{N. Bastian,
          \inst{1,2,3}
          \and
          S. Kamann\inst{3}
          \and
          F. Niederhofer\inst{4}
                    \and
          S. Saracino\inst{3,5}}

   \institute{Donostia International Physics Center (DIPC), Paseo Manuel de Lardizabal, 4, 20018, Donostia-San Sebasti\'an, Guipuzkoa, Spain\\
              \email{nate.bastian@dipc.org}
         \and
             IKERBASQUE, Basque Foundation for Science, 48013, Bilbao, Spain
             \and
             Astrophysics Research Institute, Liverpool John Moores University, 146 Brownlow Hill, Liverpool L3 5RF, UK
             \and
             Leibniz-Institut für Astrophysik Potsdam, An der Sternwarte 16, D-14482, Potsdam, Germany
             \and
             INAF – Osservatorio Astrofisico di Arcetri, Largo E. Fermi 5, 50125 Firenze, Italy
             }

   \date{Received March 31, 2025; accepted April 1, 2025}

 
  \abstract
{A number of theories have been put forward to explain the bi-modal stellar rotational distribution observed in young massive clusters.  These include stellar mergers and interactions induced in binary systems, and the role of angular momentum transfer between a star and its circumstellar disk in its early evolution.  Each theory predicts unique rotation distributions in various locations of the colour-magnitude diagram.    Specifically, the stellar merger hypothesis posits that the upper end of the main sequence will host a significant number of slowly rotating merger products, essentially that the blue straggler stars are an extension of the blue main sequence.  In the present work, we use observations of three massive ($\sim10^5$~\msun) young ($100-300$~Myr) clusters in the Large Magellanic Cloud using a combination of HST photometry and VLT/MUSE spectroscopy.  We show that in all three clusters, these bright blue stars have stellar rotational distributions that differ significantly from that measured on the blue main sequence.  We conclude that stellar mergers do not play a significant role in the formation of the split main sequence/bi-modal rotational distribution.  As a corollary, we show that blue straggler stars in these YMCs display a wide range of rotational velocities.} 

   \keywords{giant planet formation --
                $\kappa$-mechanism --
                stability of gas spheres
               }

   \maketitle
%

\section{Introduction}

The role of stellar rotation in determining the observed features in the colour-magnitude diagram of young and intermediate age clusters is now well established.  Clusters younger than $500-600$~Myr show split main sequences (in near-UV and/or blue bands - for a recent review see \citealt[]{li_review24}) that are due to a bi-modal stellar rotational distribution \citep[e.g.,][]{kamann23}. Stars on the blue main sequence (bMS) are slow rotators while the red main sequence (rMS) is populated by rapidly rotating stars. Additionally, clusters with ages up to $\sim2$~Gyr show main sequence turn-offs (MSTOs) that are significantly broader compared to expectations of simple stellar populations, which are also caused by the spread in stellar rotation rates \citep[e.g.,][]{dupree17, kamann20}.  Beyond $\sim2$~Gyr the extended MSTOs disappears abruptly as the turn-off mass drops below the threshold where stars become magnetically braked, hence all stars become slow rotators \citep[][]{martocchia18, georgy19}.

The origin of the bi-modal rotational distribution, however, is still uncertain.  To date, three hypotheses have been put forward in the literature.  The first is that tidal interactions within binary systems cause the stars to become tidally locked and subsequently slow down their rotation \citep[][]{dantona17}.  In this scenario, the slowly rotating stars (i.e., the blue main sequence - bMS) would be made up of binary stars that have been braked through binary interactions. In this scenario, we would expect the bMS to have a much higher binary fraction than the red main sequence, in conflict with observations that have shown that they are similar \citep[][]{Kamann2021}, see Appendix~A.

The second scenario is also based on binary stars. \citet[][]{wang22} suggest that stellar mergers can result in a merger product that is a slowly rotating star.  They suggest that the mergers happen mostly during the pre-main sequence (PMS) or shortly thereafter, however, the process can continue at a lower rate during the subsequent few tens or even hundreds of Myr.  Here, the bMS is formed due to two effects.  The first is that the stars are slowly rotating (hence hotter on average for a given luminosity) and the second is that the stars are rejuvenated, i.e., the merger product resets at the zero-age main sequence.  
It should be noted, however, that a key assumption of this model is that all/most merger products result in a slowly rotating star.  This is not a given, as theoretically, depending on the orientation of the orbit/collision, the merger product may result in a rapid rotator \citep[e.g.,][]{demink13}.  Simulations of merging stars have suggested that the merger product may lose a large amount of angular momentum during a "bloated phase" directly after the merger \citep[][]{schneider19}.  

The final scenario, put forward by \citet[][]{bastian20}, is that the rotation rate of a star is dictated by whether a star has been able to retain its circumstellar disk throughout its PMS phase (slow rotators), or whether the disk gets disrupted due to X-ray/UV photoionisation or dynamical encounters (fast rotators).  The scenario is based on the idea that circumstellar disks transport angular momentum away from the collapsing PMS star, so that stars that retain their disks become slow rotators, but that stars that lose their disks become fast rotators.  This scenario is best tested in very young clusters whose stars may still retain their disks.  The prediction is that there should be a strong anti-correlation between the presence of a disk and the rotation rate of the stars.  \citet[][]{bu25} have reported this trend in the young Galactic star-forming region, NGC~2264. Further studies with larger samples of stars in more regions are required to confirm this initial findings.

In the binary interaction model (the first scenario discussed above), the expectation is that the fraction of binaries amongst the bMS stars should be near unity, due to the fact that in order for a star to become a slow rotator a close and tidally interacting companion must be present. However, spectroscopic observations of bMS and red MS (rMS) stars have not found evidence for an elevated binary fraction amongst bMS stars \citep[][]{Kamann2021, he23, bu24}. 

\begin{figure}
\centering
\includegraphics[width=.5\textwidth]{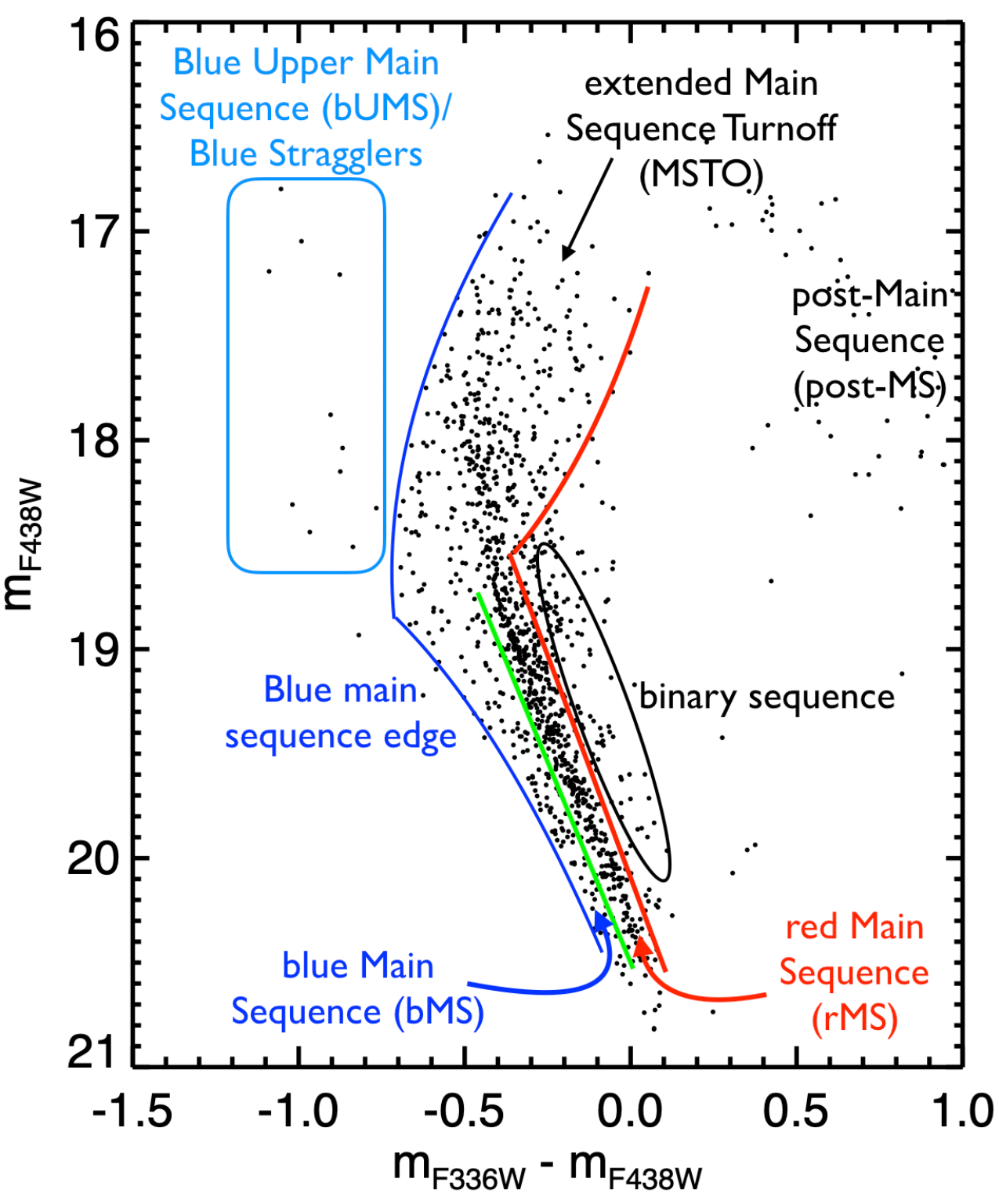}
\caption{Observed CMD of NGC~1866 with specific regions and features highlighted.}
\label{fig:cmd1}
\end{figure}

In the binary merger model (second scenario above), which is the focus of the current study, the blue straggler stars located bluewards of the extended Main eMSTO are of particular relevance. \citet{wang22} considered them as a natural extension of the blue main sequence, hence making the implicit assumption that they are merger products. We note that while detailed studies of blue stragglers at high stellar masses are still scarce, studies of low mass stars in old star clusters have shown that many blue stragglers form via stable mass transfer rather than mergers \citep[e.g.,][]{geller2011, gosnell2019}
\footnote{  However, as the bUMS stars would be gainers in mass transfer, which is independent of their evolutionary status, such stars would also be expected to be on the bMS. This means that we would expect the rotational distributions to still be similar between the bUMS and bMS under the assumption that the same mechanism (mass transfer or mergers) causes both populations.}. In this work, we investigate the validity of the assumption that blue stragglers form an extension of the blue main sequence (i.e., a blue upper main sequence), or whether they are a separate phenomenon to the bMS. For clarity, we highlight specific regions discussed in the current work in Fig.~\ref{fig:cmd1}.


The binary merger model can be tested directly by comparing the rotation rates (\vsini) of stars on the bUMS with those observed on the bMS.  In the \citet[][]{wang22} scenario, the rotation rates should be statistically the same, given that the same process is responsible for both populations.  If, however, the blue stragglers and the bMS are formed by different processes, then the rotation distribution may be very different.

In the present work we use a combination of HST photometry and VLT/MUSE spectroscopy of large samples of stars in three young clusters in the Large Magellanic Cloud (LMC), namely NGC~1850 ($90$~Myr,  2\,184 stars), NGC~1866 ($200$~Myr, 2\,531 stars) and NGC~1856 (~$300$~Myr, 4\,406 stars) to locate stars in the colour-magnitude diagram and measure their rotation rates.  The data are presented in \S~\ref{sec:data} and we discuss the results and their implications in \S~\ref{sec:results}.

\section{Data}
\label{sec:data}

For the present work, we use two independent datasets, both of which have been presented in detail in other works.

Firstly, we used the published catalogues of HST photometry for all three clusters from \citet[][]{niederhofer24}, which include the F336W, F438W and F814W bands, along with proper motion (PM) measurements. Proper motions were used to separate stars belonging to the clusters from the surrounding fields stars.  The NGC~1856 data have been corrected for differential extinction, while no such corrections were needed for, or applied to, NGC~1850 and NGC~1866.  We note that the extinction correction has a negligible impact on the results of this study.

NGC~1850 has a projected nearby neighbour, NGC~1850b, whose young age \citep[$<10$~Myr][]{gilmozzi94} results in its main sequence stars overlapping with the bUMS/blue straggler stars of NGC~1850.  Fortunately, the two clusters have PMs which can be used to separate the two populations.  We applied a PM cut of $\mu_{\alpha} cos \delta > 1.9$ to select NGC~1850 stars along with a spatial cut of RA $< 77.172$.  No such complications were seen in the NGC~1866 and NGC~1856 data, as these clusters are clearly separated from their surroundings.

Secondly, we used VLT/MUSE integral field spectroscopy in order to measure the rotation rates (\vsini) of stars in all three clusters.  Measurements for stars in NGC~1850 were taken from \citet[][]{kamann23} while those for NGC~1866 and NGC~1856 were taken from \citet[][]{kamann25}.  We refer the interested reader to those works for details of the observations, spectral extraction and data analysis.  Stellar rotation rates were measured for each star following the procedures presented in 
\citet[][]{Kamann2021} and \citet[][]{kamann23}.

\begin{figure*}    
\centering
\includegraphics[width=.95\textwidth]{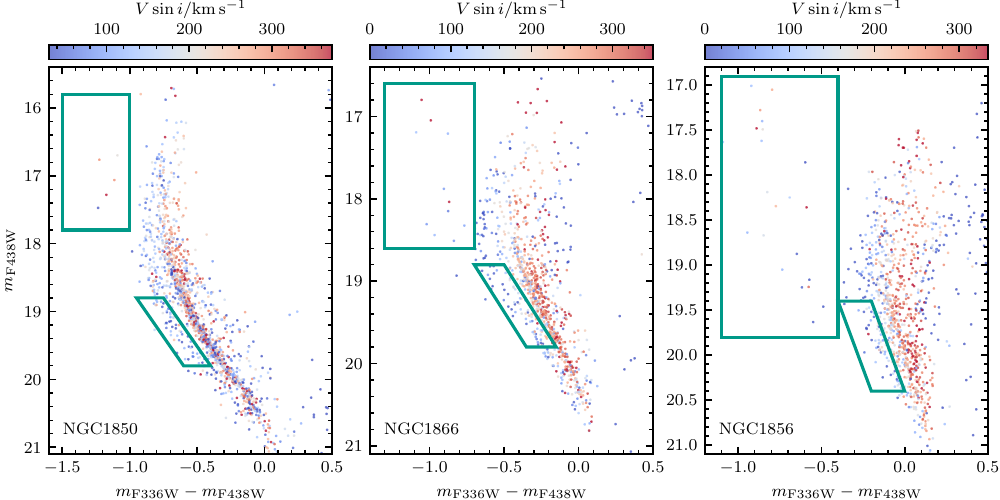}

\caption{Observed CMDs of NGC~1850 (left), NGC~1866 (middle), NGC~1856 (right).  Only stars that were considered cluster members after proper motion and radial velocity cuts are shown.  Boxes highlight the regions in each cluster that were selected for blue upper main sequence stars (upper rectangles) and blue main sequence stars (lower parallelograms).}
\label{fig:cmd_tot}
\end{figure*}

In order to make a quantitative comparison between the rotation rate distribution of stars on the bUMS and on the blue main sequence (bMS), we selected stars in the regions shown in Fig.~\ref{fig:cmd_tot}.  The bMS selection region was chosen based on the split MS.  The bUMS region was selected effectively to select classical blue straggler stars (i.e., those stars that are brighter and bluer than the main sequence turn-off region of each cluster).

\begin{figure}
\centering
\includegraphics[width=.5\textwidth]{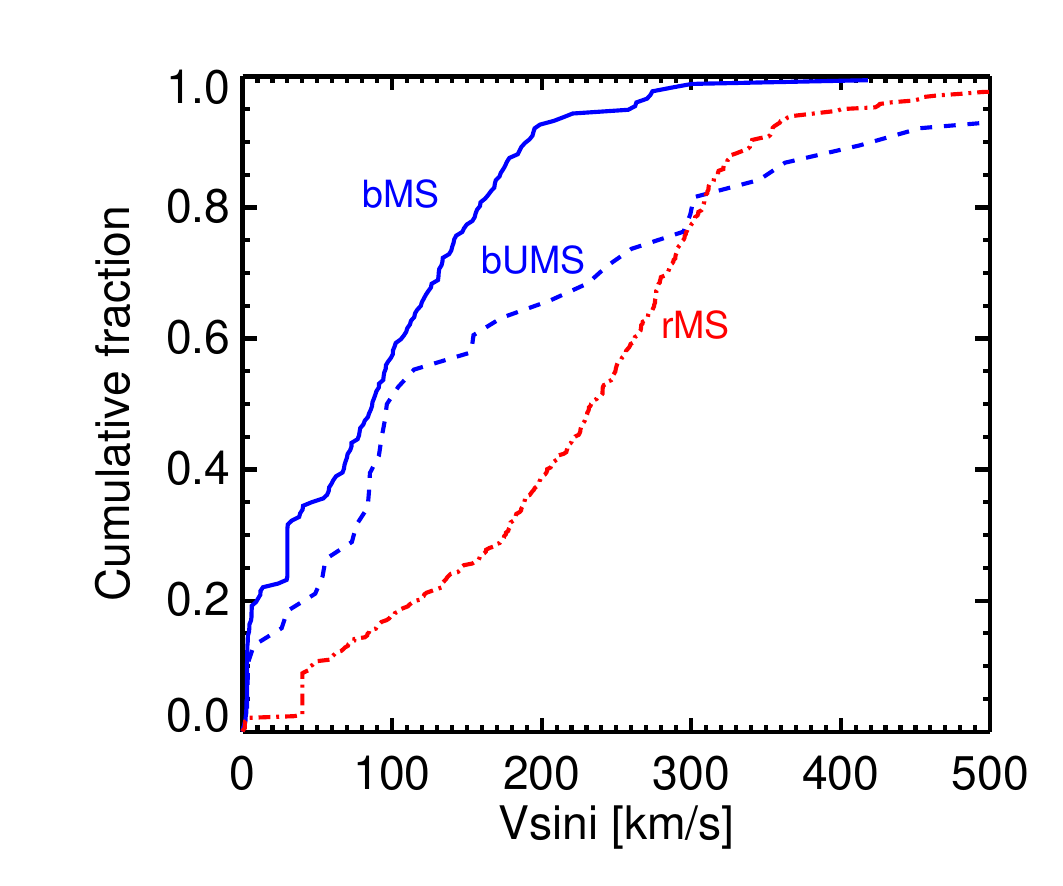}
\caption{The cumulative \vsini\ distribution of the combined bMS (blue solid line) and bUMS (blue dashed line) from all clusters.  A KS test reveals that the two populations differ significantly (p$=0.014$).  We also show the \vsini\ distribution of the rMS (red dot-dashed line) of the three clusters combined, selected from stars at similar magnitudes as the bMS.}
\label{fig:cumulative}
\end{figure}

\section{Results and Discussion}
\label{sec:results}

\citet[][]{wang22} highlighted that a significant difference in the predictions between the binary merger model and the disk evolution model can be found in stars on the bUMS (see Fig.~\ref{fig:cmd1}).  In the binary merger model, these stars are effectively the more massive counterparts to the blue main sequence stars (see Fig.~3 and Supplementary Fig.~14 from \citealt[]{wang22}).  In the model, all stars on the bMS and the bUMS are merger products.  As such, they should display the same rotational velocity distributions. Specifically, both populations are expected to be made up of slow rotators.  

Alternatively, in the disk evolution model, the bMS is made up of stars that have been able to retain their circumstellar disks during their early evolution and have been braked to low \vsini\ values.  The bUMS is an unrelated phenomenon, effectively just made up of blue straggler stars, which indeed are likely to be merger products or mass transfer systems.  In this case, we might expect that the two populations (bMS and bUMS) would have different \vsini\ distributions.

We use the datasets for the three clusters to search for such differences/similarities between the bUMS and the bMS by comparing the median, mean and standard deviation of their \vsini\ distributions.  The results are shown in Table~\ref{tab:results}.  As can be seen visually from Fig.~\ref{fig:cmd_tot}, the rotational distribution in the bUMS and the bMS are very different, with bUMS having a relatively high fraction of rapid rotators, which are unseen on the bMS.

 We have also used a Kolmogorov Smirnov (KS) test to determine the probability that each of the populations was drawn from the same parent distribution. 
 These numbers are also shown in Table~\ref{tab:results}. In all three clusters we find a low probability of the two distributions (bUMS and bMS) being drawn from the same parent populations (p$<0.025$ in all clusters).  This is in direct conflict with the predictions/assumptions of the \citet{wang22} merger scenario
 
In order to further quantify the differences we have combined populations on the bUMS and the bMS for each cluster in order to improve our number statistics.   The cumulative distributions are shown in Fig.~\ref{fig:cumulative}. A KS-test finds a low (P$=0.014$) probability that they were drawn from the same parent population.   

From these tests we find that, contrary to the predictions of the \citet{wang22} binary merger scenario, the bUMS and the bMS have different \vsini\ distributions.  This suggests that the origin of the two populations is different.  We conclude that the bMS is not made up of binary merger products, and that the blue straggler population within young clusters is not related to the observed bMS.

A corollary of these results is that not all blue straggler stars in young clusters are slow rotators.  Hence, whatever the cause of blue straggler stars, be it stellar mergers or interactions/mass transfer, it does not necessarily lead to a slowly rotating star.  

Our results qualitatively agree with investigations into the spin distributions of blue stragglers in old globular clusters, where the formation of blue stragglers is widely believed to result from a mixture of mass transfer and stellar collisions. The \vsini{} distribution of blue stragglers found in Galactic globular clusters is typically characterised by a dominant population of slow rotators and a tail that extends to larger values \citep[e.g.,][]{simunic2014,mucciarelli2014,billi2023}, with the contribution of the latter decreasing with increasing cluster density \citep{ferraro2023}. The dashed line in Fig.~\ref{fig:cumulative} resembles such a distribution, although for a higher \vsini{} on average. In globular clusters, the fast rotating blue stragglers can be linked to recent interactions, because of the fast spin-down times of low-mass stars \citep[e.g.,][]{leiner2018}. However, such a link cannot be readily adopted for our targets, because of the young cluster ages and the lack of braking experienced by early-type stars. Instead, one might speculate that the fast rotators in our sample are the result of mass transfer, while the slow rotators predominantly originate from stellar collisions \citep[e.g.,][]{bailyn92}. Future studies of the binary properties of both types of blue stragglers may provide more definite conclusions on this point.

We have only tested the specific model of \citet{wang22} and not a general binary merger model.  Our results suggest that, in analogy to older clusters, bUMS/blue stragglers in younger clusters are formed through multiple physical processes, i.e., mass transfer systems and/or stellar mergers.  These processes could be included in future models.  However, we note that such gainer stars would be expected to also be found on the main sequence \footnote{While such mass transfer products would be expected to be rapid rotators, it is not entirely clear which main sequence (blue or red) they would belong to.  It depends on the balance between rotational effects and rejuvenation.}, as being a gainer is independent of a star's evolutionary status.  Hence, in this case we may still expect similar \vsini\ distributions between the bMS and bUMS.  Additionally, such models would need to predict the relative fractions of bUMS and bMS stars. Finally, we note that the estimated age distributions of the binary mergers in the \citet{wang22} model are highly sensitive to the exact shape of the isochrones adopted, which are known to vary between different stellar evolutionary models (see Appendix~\ref{app:mergers}).

\begin{table}
\centering
\caption{Properties of the rotation rate distributions in the bUMS and the bMS for each cluster.}
\begin{tabular}{c c c c}
\hline
 & median & mean & standard deviation\\
 & (km/s) & (km/s) & (km/s)\\
\hline
{\bf NGC 1850} & & & \\
blue UMS & 299.4 &   258.0 &     155.2 \\
bMS & 97.4   & 111.3  &   71.9\\
Prob$^{a}$ = 0.01 \\
\hline
{\bf NGC 1866} & & & \\
blue UMS & 83.7 &  172.7 &  189.0 \\
bMS & 29.5  &  61.6 &  74.7\\
Prob$^{a}$ = 0.018 \\

\hline
{\bf NGC 1856} & & & \\
blue UMS & 96.5 &  149.0 & 170.4 \\
bMS & 91.2  &  95.1 & 79.2\\
Prob$^{a}$ = 0.024 \\


\hline
\end{tabular}
\tablefoot{
  $^{a}$The probability of the two samples being drawn from the same parent distribution.}
\label{tab:results}
\end{table}

\begin{acknowledgements}
Chen Wang, Selma de Mink and Norbert Langer are thanked for useful discussions on the binary merger model. SK acknowledges funding from UKRI in the form of a Future Leaders Fellowship (grant no. MR/T022868/1). FN acknowledges funding from DLR grant 50 OR 2216. SS acknowledges funding from the European Union under the
grant ERC-2022-AdG, "StarDance: the non-canonical evolution of
stars in clusters", Grant Agreement 101093572, PI: E. Pancino.  
Based on observations made with ESO Telescopes at La Silla Paranal Observatory. Based on observations of the NASA/ESA Hubble Space Telescope, obtained from the data archive at the Space Telescope Science Institute. STScI is operated by the Association of Universities for Research in Astronomy, Inc. under NASA contract NAS5-26555.
\end{acknowledgements}

%
%
\bibliographystyle{aa}
\bibliography{bums}


\appendix

\section{A reanalysis of the binary fractions of the blue and red main sequences.}
\label{app:binaries}

In addition to the stellar merger hypothesis for the origin of the bimodal stellar rotation distribution, \citet[][]{dantona17} have suggested that tidal forces between binary components could slow down rapid rotators.  The authors suggest that if all/most stars in clusters are born as fast rotators, tidal locking may slow down a significant fraction of these stars, leading to the blue Main-Sequence (bMS).  A prediction of this model is that all/most of the bMS stars should be in binary systems, specifically in close binary systems where tidal forces can lock the stars into a synchronous orbit, effectively slowing each star down. \citet[][]{Kamann2021} tested this scenario by spectroscopically measuring the binary fraction through radial velocity variations of both the bMS and rMS in the young massive cluster, NGC~1850.  They found that both sequences had a low (and statistically identical) binary fraction ($\sim5$\%), in conflict with predictions of the tidal forces scenario.

Recently, \citet[][]{muratore24} have revisited this problem by estimating the binary fractions in the bMS and rMS in three young massive clusters in the LMC.  These authors used multi-colour HST photometry, in order to identify binaries in certain filter combinations and then associate them with the different populations, using other combinations.  The method is described in detail in \citet[][]{milone20}.

\citet[][]{muratore24} report a much higher binary fraction amongst bMS stars ($23\pm4$\%) relative to rMS stars ($5\pm4$\%) in NGC~1850, and a similar trend in the other two clusters.  This result broadly supports the tidal locking scenario, however, it is in conflict with the spectroscopic binary estimates \citep[][]{Kamann2021}.  The authors offer a possible explanation, that tidal locking may be more efficient than previously thought, essentially binary stars whose orbital separation is large enough to avoid spectroscopic detection might be able to still tidally lock (i.e., that soft-binaries might be able to tidally lock).  While this is a possibility, it remains unclear why the two populations (rMS and bMS) would have similar hard binary populations.

In this section, we re-evaluate the \citep[][]{muratore24} method and subsequently their results using synthetic clusters.  Effectively, the method adopted is to look at a CMD (specifically m$_{\rm F275W}$ - m$_{\rm F336W}$ vs. m$_{\rm F336W}$) where the bMS and rMS overlap, in order to select high mass ratio binaries.  By then looking at the position of the binary systems in a different CMD projection (namely, m$_{\rm F336W}$ - m$_{\rm F18W}$ vs. m$_{\rm F336W}$), where the two populations are separated, the relative binary fractions can be inferred by looking at the resulting distribution compared to equal mass binary sequences. This is done by looking at the binary systems positions in colour space (after verticalization of the main sequence) in combination with simulated populations.  Effectively, the cumulative colour distribution of the observations is compared to that of synthetic cluster populations with different input parameters (binary fractions of the red and blue populations, fraction of red/blue MS stars, mass distributions, etc) in order to select the best fitting model. 

A potential cause for concern in the analysis of \citet{muratore24} is the relatively low number of N$\lesssim$50 binaries in their photometric selection boxes.  When stochastically sampling N data points from two distributions, with probabilities of $F_{\rm bin,red}$ and $F_{\rm bin,blue}=1-F_{\rm bin,red}$, the number of actually drawn rMS and bMS binaries will be subject to Poisson noise.  This limits the precision with which $F_{\rm bin,red}$ can be recovered from the numbers drawn.  If we adopt $F_{\rm bin,red}$ and N=38, based on Fig.~3 in \citet{muratore24}, we find that $F_{\rm bin,red}$ can be recovered to an accuracy of 8\%, which is already higher than the uncertainties provided for NGC~1850 in \citet{muratore24}. Contrary to this simple estimate, for the real data, we cannot tell for each individual binary whether it belongs to the bMS or the rMS sample, because the two sequences overlap. This will likely substantially increase the uncertainties compared to 8\% stated above. We note that \citet{muratore24} used mock samples containing about 10$\times$ the number of stars as their real samples to estimate their uncertainties. This implies that stochastic effects play a much smaller role in the mock samples than they do in the real data, potentially leading to unrealistically low uncertainties.

In order to directly test the method, we created a number of synthetic cluster datasets with the non-rotating MIST isochrones at LMC metallicity \citep[][]{choi16} and a Kroupa IMF \citep[][]{kroupa01}.  We adopted an age of $100$~Myr \citep[e.g.,][]{Bastian2016}.  Each cluster contains two populations, a bMS (containing one third of the stars) and a rMS (two thirds of the stars), offset by $\sim0.1$~mag in colour in m$_{\rm F336W}$ - m$_{\rm F814W}$ (shifted by hand to match observations) but overlapping in m$_{\rm F275W}$ - m$_{\rm F336W}$.  We assume that the binary fraction of the red population is constant at 0.1 (F$_{\rm bin,red}$).  For the blue population, we adopt different binary fractions (F$_{\rm bin,blue}$) of 0.1, 0.2 or 0.4.  For all binaries we adopt a flat mass-ratio distribution. For each combination of parameters we create ten synthetic clusters that are stochastically sampled.

\begin{figure*}
\centering
\includegraphics[width=.45\textwidth]{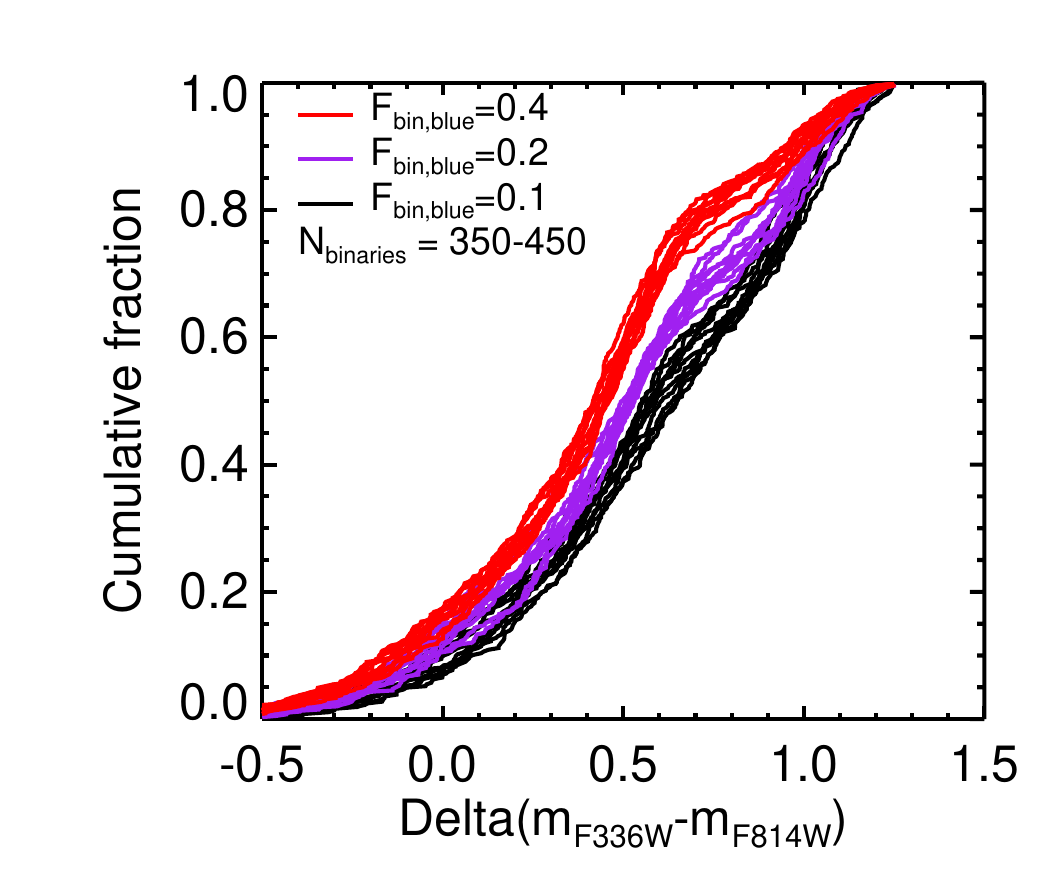}
\includegraphics[width=.45\textwidth]{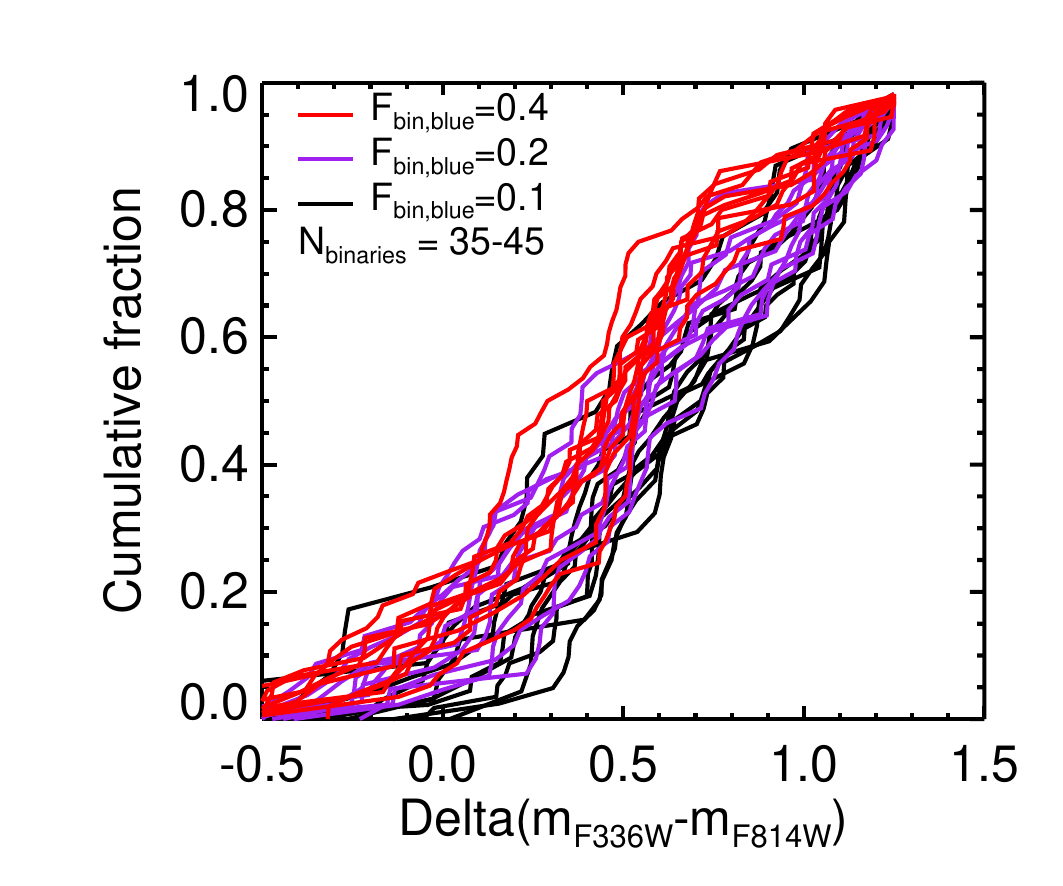}
\caption{{\bf Left panel:} The cumulative distribution of the synthetic clusters, separated by colour for the three different F$_{\rm bin,blue}$ values.  Each synthetic cluster has 350 to 450 binary systems that fall into the selection box, representing the high-number regime. {\bf Right panel:} The same as the left panel but now for synthetic clusters with only 35 to 45 binary systems in the selection box in order to be more consistent with the observed clusters.}
\label{fig:appendix_cumul1}
\end{figure*}

In a first experiment, we create stochastic clusters with $\sim10$x more stars than observed (i.e., for NGC~1850 there are 38 binary systems in the selection box used by \citet[][]{muratore24}), meaning that the synthetic clusters have 350 to 450 binary systems in the selection box.  This is meant to broadly reproduce the simulations used in \citet[][]{muratore24}.  The resulting cumulative distributions are shown in the left panel of Fig.~\ref{fig:appendix_cumul1}.

From this figure we can see that the overall method is sensitive to the binary fraction, and that the cumulative distributions are significantly different for different F$_{\rm bin,blue}$ values.  Of course, all other parameters in the model are fixed, leading to an idealized simulation, but the results are clearly very sensitive to F$_{\rm bin,blue}$. 

In a second set of simulations, we attempt to reproduce the observations more accurately, specifically the number of binary systems that fall within the selection box.  \citet[][]{muratore24} found 38 binary systems (that they used for the analysis) within their selection box.  For this second set of simulations, the synthetic clusters have the same set of parameters as previously adopted, but now they have only 35 - 45 binary systems that fall into the selection box.

The results are shown in the right panel of Fig.~\ref{fig:appendix_cumul1}.  As can be seen, in the low-N limit (and we note that NGC~1850 had the highest number of binary systems in the selection box of the three clusters studied by \citet[][]{muratore24}) the diagnostic power of this method becomes significantly worse.  In this case, even synthetic clusters with radically different F$_{\rm bin,blue}$ values overlap, meaning that no significant constraints can be put on the binary fractions of the two populations.  Again, we note that in these idealised simulations, all other parameters are known and kept fixed, limiting any additional uncertainties.

We conclude that the method adopted by \citet[][]{muratore24} is potentially very powerful in the high-N regime.  However, in the space occupied by young massive clusters, the method is not able to provide useful constraints.  Hence, we conclude that the uncertainties cited by \citet[][]{muratore24} are underestimated due to the lack of inclusion of accurate number statistics.  This is likely the reason  for the discrepancy between the estimated binary fractions between that work and \citet[][]{Kamann2021}.

\section{Magnitude dependent merger histories}
\label{app:mergers}

\citet{wang22} derive the merger histories of binaries in their sample of clusters by fitting the position of stars along the bMS with rejuvenated stellar models.  Under the assumption that each star is a merger product (in this case from an equal mass merger) the position of the star between the current isochrone that fits the bMS (i.e., the age of the cluster) and the zero-age main sequence can be used to determine when the merger happened. This is based on the idea that a merger would reset the main sequence evolution of a star.

\citet{wang22} note that towards the main sequence turn-off more stars are to the blue of the best fitting bMS isochrone compared to fainter magnitudes.  The authors use this to conclude that the binary merger history is magnitude (i.e., mass) dependent.  

An alternative explanation is that uncertainties in the models lead to small offsets between the observations and the isochrones.  This is to be expected to some extent, given the uncertainties and parameterizations within stellar evolutionary models (especially when rotation is considered - see e.g., \citealt[][]{ekstrom18}).

\begin{figure}
\centering
\includegraphics[width=.5\textwidth]{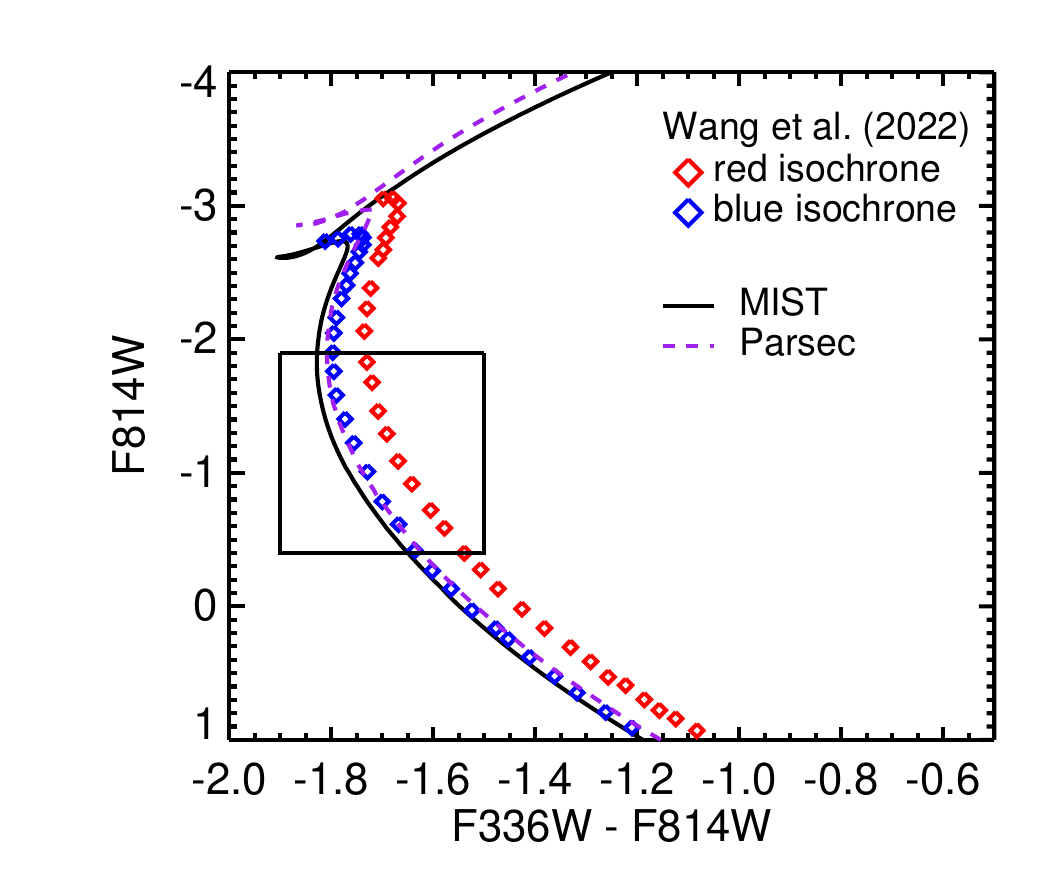}
\caption{A comparison of the isochrones from three different stellar evolutionary codes.}
\label{fig:appendix_isochrones}
\end{figure}

As an example, we can look at NGC~330 as studied in Fig.~3 in \citet{wang22}.  At a magnitude of m$_{\rm F814W} \sim 18.5$ we see a cloud of stars that lie $m_{\rm F336W} - m_{\rm F814W} \sim 0.05-0.1$~mag to the blue of the adopted bMS isochrone (translating into the box in Fig.~\ref{fig:appendix_isochrones}).

To get a feel on the level of model uncertainties, we compared the adopted \citet{wang22} isochrones to those of other groups.  Specifically, we adopt the MIST \citep{choi16} and Parsec models \citep{parsec12}.  We remove the extinction and distance modulus applied to the isochrones used in \citet{wang22} and adopt the same metallicity for all models (namely Z$_{\rm SMC} = 0.00218$) and used the non-rotating models.  A comparison of the isochrones are shown in Fig.~\ref{fig:appendix_isochrones}.

While all three models show a very similar shape and at fainter magnitudes the agreement between the models is excellent.  However, we do see differences at brighter magnitudes.  Specifically, in the box highlighted in Fig. ~\ref{fig:appendix_isochrones} (which is where the main offset between the models and data in \citet{wang22} is found) the Parsec and MIST models are not as steep as the \citet{wang22} models, leading to bluer models.  Within the region of interest, the offset is $m_{\rm F336W} - m_{\rm F814W}\sim0.05$~mag.  This is similar to the reported offset between the \citet{wang22} isochrones and the observed stars, highlighting that caution is necessary when comparing small offsets between data and isochrones.

While it is impossible to make any statements about which model is best, this comparison demonstrates that caution is needed when comparing the predictions of stellar isochrones with actual data, as offsets may not have a physical origin.

\end{document}